*AI generated annotations for Breast, Brain, Liver, Lungs and Prostate cancer collections in National Cancer Institute Imaging Data Commons*

Gowtham Krishnan Murugesan[1], Diana McCrumb[1], Rahul Soni[1], Jithendra Kumar[1], Leonard Nuernberg[4], Linmin Pei[3], Ulrike Wagner[3], Sutton Granger[2], Andrey Y. Fedorov[4], Stephen Moore[1], Jeff Van Oss[1]

**Affiliations**
1. BAMF Health, Grand Rapids, MI, USA
2. National Institute of Health, Bethesda, MD, USA
3. Frederick National Laboratory for Cancer Research, Frederick, MD, USA
4. Brigham and Women's Hospital and Harvard Medical School, Boston, MA, USA
corresponding author(s): Gowtham Krishnan Murugesan (gowtham.murugesan@bamfhealth.com)

## Abstract

AI in Medical Imaging project aims to enhance the National Cancer Institute's (NCI) Image Data Commons (IDC) by developing nnU-Net models and providing AI-assisted segmentations for cancer radiology images. We created high-quality, AI-annotated imaging datasets for 11 IDC collections. These datasets include images from various modalities, such as computed tomography (CT) and magnetic resonance imaging (MRI), covering the lungs, breast, brain, kidneys, prostate, and liver. The nnU-Net models were trained using open-source datasets. A portion of the AI-generated annotations was reviewed and corrected by radiologists. Both the AI and radiologist annotations were encoded in compliance with the the Digital Imaging and Communications in Medicine (DICOM) standard, ensuring seamless integration into the IDC collections. All models, images, and annotations are publicly accessible, facilitating further research and development in cancer imaging. This work supports the advancement of imaging tools and algorithms by providing comprehensive and accurate annotated datasets.

## Background & Summary

Advances in AI for medical imaging, especially deep learning (DL), have led to significant progress in tumor and organ segmentation models. Developing Reliable AI models for cancer imaging require large datasets with high-quality annotations, but manual annotation is labor-intensive. The availability of reliable annotations is critical to supervise machine learning-based algorithms for several downstream clinical tasks. However, many image collections in IDC lack reliable annotations of tumors, organs, or tissues. To address this, we developed nn-UNet models to generate accurate tumor segmentations from publicly available imaging data, aiding research and downstream AI model development.

In our previous work[1], we enriched 1925 radiological images from 11 distinct IDC collections with annotations. In this project, we aim to further enhance various other IDC collections by developing nn-UNet models for AI-assisted segmentations. We continue to provide both models and segmentations for a selected subset of IDC cancer datasets. We developed state-of-the-art nnUNet models for organ and lesion segmentations using publicly available datasets. This effort resulted in an AI-annotated imaging dataset that includes tissues, organs, and cancers 11 IDC image collections. These collections feature images from various modalities, including computed tomography (CT) and magnetic resonance imaging (MRI), and cover several body parts such as the chest, breast, kidneys, prostate, and liver. To ensure the accuracy of the AI models, a portion of the AI-generated annotations was reviewed and corrected by a board-certified radiologist. Both the AI and radiologist annotations were encoded in compliance with the DICOM standard, facilitating seamless integration into the IDC collections as third-party analysis sets. All models, images, and annotations are publicly accessible, supporting further research and development in cancer imaging.

## Methods

In this project, we developed nnU-Net models to generate AI segmentations for six distinct tasks: brain, breast, lung, liver, prostate, and kidney organ and tumor segmentation. These tasks utilize 11 unique IDC collections, as outlined in Table 1. New models were trained for four of these tasks: brain, breast, liver, and lung using publicly available datasets detailed in Table 1. For the kidney and prostate segmentation tasks, we employed models developed as part of previous work to enrich additional IDC collections with AI assisted annotations.

**Model Training and Data Processing**

**Brain Tumor Segmentation**: We used the BRATS 2021[2] dataset, which includes 1251 paired T1, T1 post-contrast, T2, and FLAIR images, to segment edema, tumor core, and enhancing tumor regions to train an nnUNet model. To enhance model generalizability, we permuted the input MRI sequence order during training.

**Breast, Fibroglandular Tissue, and Tumor Segmentation**: The nnUNet model was trained using 489 post-contrast T2 images from the Duke-Breast-Cancer-MRI-Supplement-v3[3] dataset. For breast tumor segmentation, an additional nnUNet model was trained using 98 post-contrast T2 MR images from the TCIA-ISPY1-Tumor-SEG-Radiomics[4,5] dataset. Tumor segmentations from this dataset includes both enhancing and non-enhancing tumor regions, defining the structural tumor volume (STV). The output from each model was combined to create a single segmentation output with three labels: breast, fibroglandular tissue (FGT), and tumor.

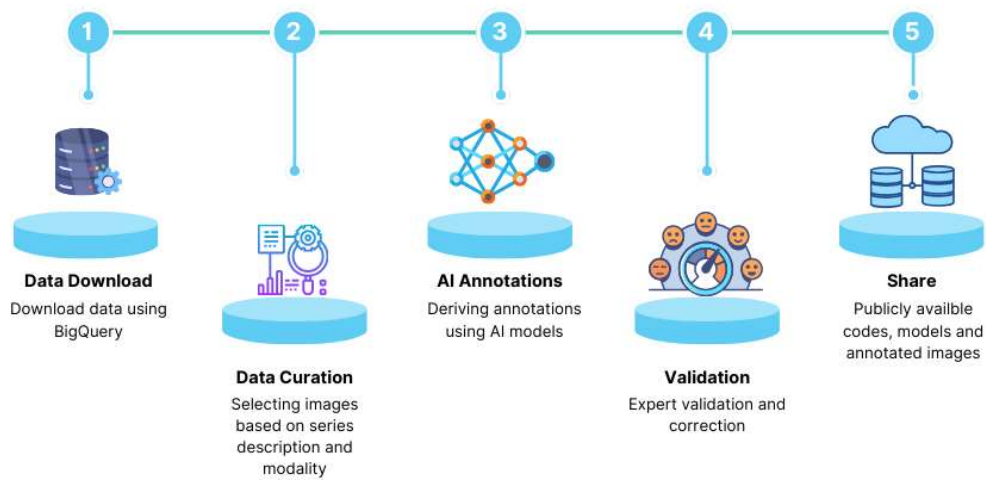

*Figure 1: AIMI workflow: Describes overall workflow from Data Download, Data Curation, Annotations, Validation and Sharing of AI annotation dataset and model weights*

*Table 1: IDC collections Enriched with AI-assisted annotations: Summarizes IDC collections, detailing the number of studies, imaging modalities of interest (MOI), curated studies enriched with AI annotations, specific annotations, training data for each task, and the number of studies validated by radiologists.*

| Task | IDC Collections Enriched (Studies) | MOI | Curated Images with MOI | Segments | Training Data/ Model (Dataset Size) | Images validated by Radiologists |
|---|---|---|---|---|---|---|
| **Brain-MR** | UPENN-GBM (630)[6] | MR (T1, T2, FLAIR, T1c) | 541 | Whole Tumor, Enhancing Tumor, Non-Enhancing Tumor | Brats2021[2] (1251) | 45 |
| **Breast-MR** | Duke-Breast-Cancer-MRI (922)[7] | MR (T1 post contrast) | 805 | Breast, FGT, and Tumor | TCIA-ISPY1-Tumor-SEG-Radiomics[3] (98), Duke-Breast-Cancer-MRI-Supplement-v3[4] (489) | 92 |
| **Kidneys-CT** | TCGA-KICH (12)[16], TCGA-KIRP (23)[17], and CPTAC-CCRCC (57)[18] | CT (Post contrast) | 64 | Kidneys, Cysts, Tumors | Kidney-CT12 BAMF-AIMI model | 7 |
| **Lung CT** | QIN Lung CT (47)[21], SPIE AAPM Lung CT Challenge (70)[22], and National Lung Screening Trial (1000)[23] | CT | 1137 | Lungs nodules | NSCLC Radiomics11 (416), LIDC_IDRI19 (883), | 114 |
| **Liver-CT** | HCC_TACE_Seg (105)[24], Colorectal-Liver-Metastases(197)[25] | CT | 515 | Liver, Tumors | Medical Decathlon Dataset8 (131), LiTS 20177 (131) | 52 |
| **Prostate-MR** | Prostate MRI US-Biopsy (842)[26] | MR (T2) | 817 | Prostate | Prostate-MR13 BAMF AIMI model | 81 |

DICOM images downloaded from IDC were converted into the Neuroimaging Informatics Technology Initiative (NIfTI) format, then all the selected MR contrasts images were reoriented to RAS, N4 Bias Corrected, registered to T1c, skull stripped and then registered to MNI[23] template before deriving AI annotations of tumor components using Brain-MR model. The AI derived annotations were post processed for inverse reorientation, inverse registration with T1c to get them back to each of MR contrast images space. The AI derived annotation in MNI space with different permutations of input order was compared with automatic annotations in UPENN-GBM dataset (Supplementary Table.1)

**Liver and Tumor Segmentation**: CT images from the LiTS 2017[20] and Medical Decathlon datasets[19] were used to train the nnUNet model for liver and tumor segmentation. We utilized selected TotalSegmentator[24] outputs to develop anatomically informed model. The liver-CT model was trained to predict liver and liver tumors, as well as other abdominal organs including the duodenum, gallbladder, intestines, kidneys, lungs, pancreas, and spleen.

DICOM images downloaded from IDC were converted into NIfTI images and used as input to the Breast-MR model to generate AI Annotations. Predictions outside breast were removed using connected component analysis.

**Lung and Nodules Segmentation**: The model for lung and nodules segmentation was trained using 883 CT images from the LIDC-IDRI[25] dataset and 416 CT images from the NSCLC Radiomics[15] dataset, each annotated for lung lesions and nodules. Annotations for the lung regions in the training dataset were generated by TotalSegmentator [24].

DICOM images downloaded from IDC were converted into NIfTI images and used as input to the Lung-CT model to generate AI Annotations. Specifically, we selected the same subset of the NLST as specified by Krishnaswamy et al[26]. Predictions outside Lungs were removed using connected component analysis. Nodules not between the size 3mm-30mm are removed from the predictions.

**Kidney and Prostate Segmentation**: For kidney tumors and cysts, and prostate segmentation, we utilized the kidney-CT[11] and prostate-MR[22] models from the BAMF AIMI project.

Details on the training data, input image types, and output segmentations for each of the models are provided in Table 1. Comprehensive information on the training, pre-processing, and post-processing procedures is available on GitHub repositories (Table.5)

**Data Curation**
For each task, source radiological images from publicly available NCI IDC collections were selected with BigQuery commands and then downloaded, with code made available in GitHub repositories (Table.5). These images were filtered to match the modality of interest requirements (Table.1) for each specific task based on the model inputs. Given the large size of the National Lung Screening Trial, 26408 cases, a subset of 1042 were selected using the query specified by Krishnaswamy et al[26] to include only CT scans of subjects who were clinically confirmed positive for lung cancer. Detailed information on the modalities of interest for each task and the number of curated images is listed in Table 1.

**Quality Assessment**
To ensure the quality of AI-generated annotations, approximately 10% of these annotations were evaluated by radiologists (Table.1). Quality metrics such as Dice coefficient, normalized surface distance (NSD), and detection accuracy were reported. We have provided code for reproducibility calculating quality metrics and enabling the downloading of data from IDC (Table.5).

To control project costs, the radiologists agreed to review and correct AI generated segmentations based on an estimation of the time required, not the actual time spent. The Radiologists used 3DSlicer to load the images and segmentations. For all tasks except Brain, the scans were loaded from DICOM format. For the Brain task, the scans were loaded from the *NIfTI* formatted files after they had been pre-processed for skull stripping and co-registration. The AI segmentation files were loaded from NRRD formatted files. The radiologists reviewed and edited the segmentations as needed to ensure accuracy and quality. After making corrections, they returned the updated segmentations in NRRD format. These corrected segmentations were subsequently converted to DICOM-SEG format for inclusion in the IDC. For the Brain task, segmentations were transformed back to the coordinate space of each of the original scans before conversion to DICOM-SEG.

The overall workflow of our analysis is illustrated in Figure 1 and representation of AI annotations generated is shown in Figure.2.

## Data Records
The reviewers scoring and comments, as well as DICOM Segmentation objects for the AI predictions and reviewer's corrections are available in Zenodo[64] (https://zenodo.org/records/13244892).

Each zip file in the collection correlates to a specific segmentation task. The common folder structure is:
- *ai-segmentations-dcm* This directory contains the AI model predictions in DICOM-SEG format for all analyzed IDC collection files.
- *qa-segmentations-dcm* This directory contains manual corrected segmentation files, based on the AI prediction, in DICOM-SEG format. Only a fraction, ~10%, of the AI predictions were corrected. Corrections to the AI model segmentations were performed by a radiologist.
- *qa-results.csv* CSV file linking the study/series UIDs with the ai segmentation file, radiologist corrected segmentation file, radiologist ratings of AI performance. Reviewer Likert scores and review comments for the segmentations are also included in this file. (Table 2 and 3)

*Table 2: Common columns and their descriptions for qa-results.csv files. Some columns in qa-results.csv are answers to questions posed to the reviewers. These questions were specific to the task the csv file refers to. For example, "Was the AI predicted Cyst label accurate?" was asked for the kidney task, but variations asking about each segment were asked about every task*

| Column | Description |
| --- | --- |
| **Collection** | The name of the IDC collection for this case |
| **PatientID** | PatientID in DICOM metadata of scan. Also called Case ID in the IDC |
| **StudyInstanceUID** | StudyInstanceUID in the DICOM metadata of the scan |
| **SeriesInstanceUID** | SeriesInstanceUID in the DICOM metadata of the scan |
| **Validation** | true/false if this scan was reviewed by a radiologist |
| **Reviewer** | Coded ID of the reviewer. Radiologist IDs start with 'rad' non-expect IDs start with 'ne' |
| **AimiProjectYear** | 2023 or 2024, This work was split over two years. The main methodology difference between the two is that in 2023, a non-expert also reviewed the AI output, but a non-expert was not utilized in 2024. |
| **AISegmentation** | The filename of the AI prediction file in DICOM-seg format. This file is in the ai-segmentations-dcm folder. |
| **CorrectedSegmentation** | The filename of the reviewer corrected prediction file in DICOM-seg format. This file is in the qa-segmentations-dcm folder. If the reviewer strongly agreed with the AI for all segments, they did not provide any correction file. |
| **Was the AI predicted * label accurate?** | This column appears one for each segment in the task. The reviewer rates that segment quality on a Likert scale |
| **Do you have any comments about the AI predicted ROIs?** | Open ended question for the reviewer |
| **Do you have any comments about the findings from the study scans?** | Open ended question for the reviewer |

*Table 3: Likert Score description used by reviewers to assess the quality of the AI annotations per case in the Validation set.*

| *Likert Score* | Description |
| --- | --- |
| **Strongly agree** | Use-as-is (i.e., clinically acceptable, and could be used for treatment without change) |
| **Agree** | Minor edits that are not necessary. Stylistic differences, but not clinically important. The current segmentation is acceptable. |
| **Neither agree nor disagree** | Minor edits that are necessary. Minor edits are those that the review judges can be made in less time than starting from scratch or are expected to have minimal effect on treatment outcome. |
| **Disagree** | Major edits. This category indicates that the necessary edit is required to ensure correctness, and sufficiently significant that user would prefer to start from the scratch. |
| **Strongly disagree** | Unusable. This category indicates that the quality of the automatic annotations is so bad that they are unusable. |

The DICOM segmentation files contain self-describing information in the metadata.

- Each segmentation can be linked back to the original DICOM scans from the DICOM metadata in each segmentation file. The DICOM data element SeriesInstanceUID (0020,000E) in the data element ReferencedSeriesSequence (0008,1115) refers to the original DICOM scan this segmentation was derived from.
- The SegmentSequence (0062,0002) contains metadata about each of the segments such as:
    - SegmentNumber (0062,0004) - The numerical value of pixels that comprise this segment
    - SegmentDescription (0062,0006) – A human readable description of the segment
    - SegmentAlgorithmType (0062,0008) – Either *AUTOMATIC* for AI model outputs or SEMIAUTOMATIC for manual reviewer corrections of the AI output.

The DICOM segmentations have been integrated into the IDC. From that portal it is possible to view the segmentations overlayed on the images they were derived from. The direct link to the segmentation collection is in IDC[64] (https://portal.imaging.datacommons.cancer.gov/explore/filters/?analysis_results_id=BAMF-AIMI-Annotations)
.

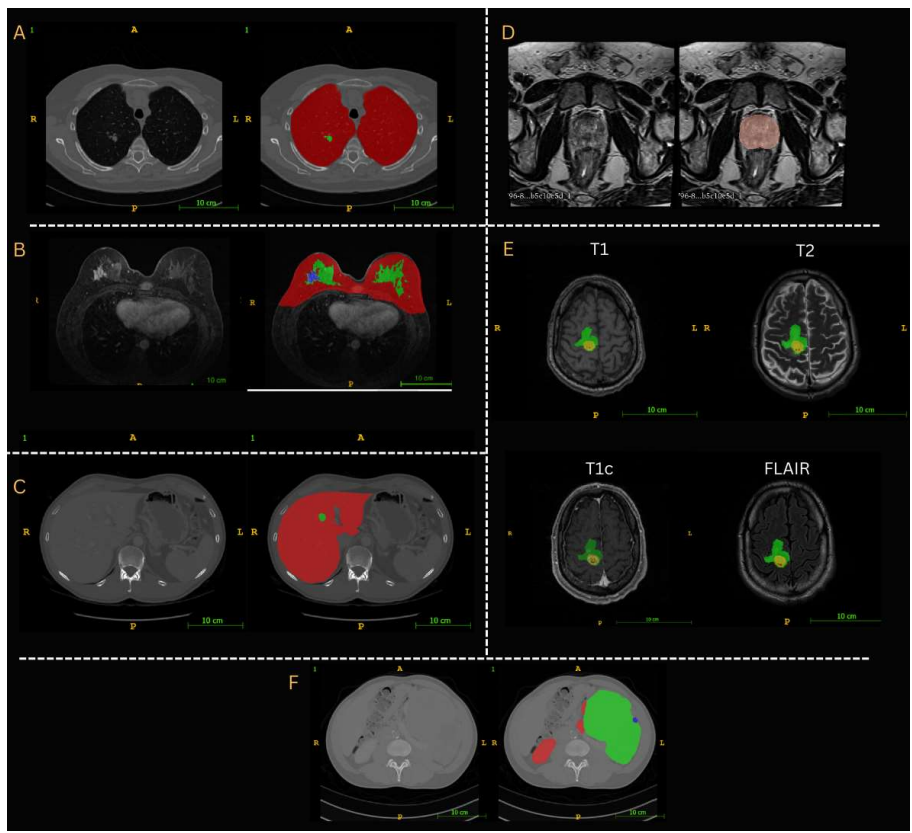

*Figure 2: AI Annotations: Representations of input images and AI annotations for A) Lungs and nodules, B) Breast, FGT and Tumor, C) Liver and Tumor, D) Prostate, E) Brain Tumor components (Edema, Enhancing and Non enhancing) and F) Kidneys, Cysts, and Tumor*

## Technical Validation

The AI models were evaluated on the following series of metrics. Some of these were only applicable to a subset of the model tasks.

- Sørensen–Dice coefficient27 (DSC): measures the similarity between volumetric segmentations, $V_A$ and $V_B$. It is twice the intersection of the volumes over the sum of the volumes.

$$DSC = \frac{2(V_A \cap V_B)}{V_A + V_B}$$

- Normalized Surface Dice28 (NSD): measures surface distance similarity. It measures the amount of the surface of a volume($S_A$) that is within a tolerance ($\tau$) of the surface of another volume ($S_B^\tau$). This is calculated for both surfaces and normalized to the total surface of the volumes. NSD tolerance level for each task were selected based on the acceptable error for the segmentation provided by Antonelli M et al.,[19]

$$NSD = \frac{(S_A \cap S_B^\tau) + (S_B \cap S_A^\tau)}{S_A + S_B}$$

- 95% Hausdorff Distance: measures surface agreement. It is the distance at which 95% of the points on Surface A have a point on Surface B less than it.

For each task, 10% of the images enriched for IDC imaging collections were evaluated and corrected by expert radiologists. The number of images evaluated, along with the Dice coefficient, 95% Hausdorff distance, and normalized surface distance for each of the model outputs for each task, are detailed in Table 4. The higher Dice scores observed in certain tasks, such as Kidneys-CT, may be attributed to experts not fully correcting the AI-generated annotations. This may reflect correction bias, as experts were refining AI-derived segmentations rather than generating them independently from scratch.

*Table 4: Quantitative Analysis: Quantitative metrics annotations for each task. NSD tolerance level for each task were selected based on the acceptable error for the segmentation provided by Antonelli et al.,* [19]

| Model | Segmentations | Dice | 95% Hausdorff Distance | NSD (tolerance(mm)) |
|---|---|---|---|---|
| **Brain-MR** | Whole Tumor | 0.98±0.07 | 6.88±0.34 | 0.98±0.04[2] |
| | Enhancing Tumor | 0.95±0.13 | 6.57±0.24 | 0.99±0.03[2] |
| | Nonenhancing Tumor | 0.97±0.08 | 0.42±1.05 | 0.97±0.09[2] |
| **Breast-MR** | Breast | 0.99±0.01 | 0.74±2.92 | 0.09±0.22[3] |
| | Fibroglandular Tissue | 0.80±0.29 | 8.75±12.92 | 1.82±4.17[2] |
| | Lesions | 0.57±0.36 | 41.44±51.78 | 9.36±13.79[2] |
| **Kidneys-CT** | Kidneys | 1.0±0.0 | 0.00±0.00 | 0.00±0.00[3] |
| | Cysts | 1.0±0.0 | 0.00±0.00 | 0.15±0.38[2] |
| | Tumors | 1.0±0.0 | 0.00±0.00 | 0.00±0.00[2] |
| **Lung-CT** | Lungs | 1.0±0.0 | 0.00±0.00 | 0.02±0.11[2] |
| | Nodules | 0.78±0.28 | 62.07±10.54 | 10.54±14.43[2] |
| **Liver-CT** | Liver | 0.99±0.02 | 2.33±7.70 | 0.29±0.95[7] |
| | Tumors | 0.80±0.35 | 19.73±38.35 | 4.38±8.70[2] |
| **Prostate-MR** | Prostate | 0.99±0.02 | 1.07±1.24 | 0.15±0.18[4] |

## Usage Notes

The AI models and datasets developed in this project are designed for broad accessibility and replication within the research community. Key aspects of their usability are outlined below:

**Model Accessibility**

The trained AI model weights are available on Zenodo, providing researchers with easy access to these resources. Specific URLs for downloading the model weights are listed in Table 5.

**Reproducibility**

To support the reproducibility of our analysis, we have provided comprehensive Jupyter notebook code on GitHub. This repository includes all necessary steps for data processing, model training, and evaluation, enabling researchers to replicate our methodology and validate our results.

**Data Availability**

The curated datasets, including both AI-generated and radiologist-corrected annotations, are accessible through the National Cancer Institute's Imaging Data Commons (IDC). This ensures that the data utilized in this study can be accessed and employed by the wider research community.

**Documentation and Support**

Detailed documentation is included within the GitHub repository, covering installation requirements, usage instructions, and troubleshooting tips. The repository also contains examples and guidelines to assist users in applying the models to their own datasets.

**Evaluation Metrics**

To ensure the quality and reliability of the AI-generated annotations, we have included evaluation metrics such as the Dice coefficient, normalized surface distance (NSD), and detection accuracy. These metrics provide quantitative measures of the model's performance and facilitate comparison in future research.

**Community Contributions**

We encourage contributions from the research community to enhance the usability and functionality of the provided tools and datasets. Users are invited to submit issues, suggest improvements, and contribute code via the GitHub repository.

By providing these resources and supporting documentation, we aim to promote the usability and reproducibility of our work, thereby facilitating advancements in medical image analysis and AI research.

## Code Availability

The AI model weights are accessible on Zenodo, and Jupyter notebook code to reproduce the analysis is available on GitHub. The models have also been released on the https://MHub.ai platform. The URLs are provided in Table 5.

*Table 5: URLs for Model weights and GitHub repositories*

| Task | Model Weights | Notebook Code |
|---|---|---|
| **Brain-MR** | https://doi.org/10.5281/zenodo.11582627 | https://github.com/bamf-health/aimi-brain-mr |
| **Breast-MR** | https://doi.org/10.5281/zenodo.11998679 | https://doi.org/10.5281/zenodo.13851641 |
| **Kidneys-CT** | https://doi.org/10.5281/zenodo.8277846 | https://doi.org/10.5281/zenodo.13851351 |
| **Lung CT** | https://doi.org/10.5281/zenodo.11582738 | https://doi.org/10.5281/zenodo.13851613 |
| **Liver-CT** | https://doi.org/10.5281/zenodo.11582728 | https://doi.org/10.5281/zenodo.13851682 |
| **Prostate-MR** | https://doi.org/10.5281/zenodo.8290093 | https://doi.org/10.5281/zenodo.13851368 |

## Acknowledgements

This project was funded, in whole or in part, with Federal funds from the National Cancer Institute, National Institutes of Health, under Contract No. 75N91019D00024, Task Order No. 75N91020F0003. The content of this publication does not necessarily reflect the views or policies of the Department of Health and Human Services, nor does mention of trade names, commercial products, or organizations imply endorsement by the U.S. Government. The results published here are based, in whole or in part, on data generated by the TCGA Research Network: http://cancergenome.nih.gov/.

## Author contributions
Gowtham Murugesan was responsible for development of the brain, breast, lungs, kidneys, and liver AI models. Gowtham also performed analysis of the AI vs QA annotations and drafted the manuscript. Diana McCrumb was responsible for the development of the prostate AI model and was the primary editor of the manuscript. Gregor Sutton reviewed and corrected the AI annotations as a radiologist. Rahul Soni, Jithendra Kumar, and Leonard Nuernberg facilitated the publication of AI models in AI model hub (https://MHub.ai). Andrey Fedorov managed dataset compatibility and integration with IDC collections. Andrey also reviewed the manuscript. Linmin Pei managed the project for the NCI and reviewed the manuscript. Keyvan Farahani, Ulrike Wagner and Stephen Moore reviewed and provided guidance for the project. Jeff Van Oss served as PI for the project and was responsible for converting files to and from DICOM, managing deliverables of code and annotations, and editing the manuscript.

## Competing interests
No conflicting interests.

*Supplementary Table1: Quantitative Analysis: Evaluation of Brain-MR model on different possible permutation of input MR contrasts (T1,T2,FLAIR,and T1c)*

|  | WT | Edema | ET | NET |
|---|---|---|---|---|
| **Dice** | 0.97+/-0 | 0.93+/-0.1 | 0.81+/-0.2 | 0.92+/-0.1 |
| **Haussdorff** | 8.82+/-11.80 | 11.68+/-10.36 | 12.34+/-11.67 | 6.03+/-9.29 |
| **Jaccard Distance** | 0.06+/-0.05 | 0.13+-0.11 | 0.28+/-0.25 | 0.13+/-0.11 |
| **FPV** | 2.11+/-2.35 | 2.20+/-3.04 | 1.49+/-3.06 | 0.94+/-0.94 |
| **FNV** | 2.40+/-3.14 | 3.35+/-3.90 | 0.54+/-1.61 | 1.04+/-1.64 |